\documentclass[twocolumn]{jpsj3}

\usepackage{color}
\usepackage{times}
\usepackage{graphicx}

\title{Kondo Effect of a Jahn-Teller Ion
Vibrating in a Cubic Anharmonic Potential}

\author{Takashi Hotta}

\inst{Department of Physics, Tokyo Metropolitan University,
Hachioji, Tokyo 192-0397, Japan}

\recdate{\today}

\abst{
We discuss the Kondo effect in a spinless two-orbital conduction electron system
coupled with anharmonic Jahn-Teller vibration
by employing a numerical renormalization group technique.
When a temperature $T$ is decreased, we encounter a plateau of $\log 3$ entropy
due to quasi-triple degeneracy of local low-energy states,
composed of vibronic ground states and the first excited state
with an excitation energy of $\Delta E$.
Around at $T$$\approx$$\Delta E$, we observe an entropy change from $\log 3$
to $\log 2$.
This $\log 2$ entropy originates from the rotational degree of freedom of
the vibronic state and it is eventually released due to the screening
by orbital moments of conduction electrons,
leading to the Kondo effect of a Jahn-Teller ion.
The Kondo temperature is explained by the effective $s$-$d$ model
with anisotropic exchange interactions.
}

\kword{Kondo effect, Dynamical Jahn-Teller effect, Vibronic state,
Cubic anharmonicity}

\begin{document}
\maketitle

%%%%%%%%%%%%%%%%%%%%%%%%%%%%%%%%%%%%%%%%%%
%   Introduction
%%%%%%%%%%%%%%%%%%%%%%%%%%%%%%%%%%%%%%%%%%
\section{Introduction}

One of recent trends in the research field of condensed matter physics
is to explore exotic magnetism and unconventional superconductivity
in heavy-electron materials.\cite{ICHE2010,SCES2013}
Concerning the emergence of heavy-electron state,
a traditional mechanism is based upon quantum criticality
induced by the competition
between the Kondo effect and Ruderman-Kittel-Kasuya-Yosida interaction.
The Kondo effect due to local magnetic moment has been well understood,
\cite{Kondo40} but it is recognized that
the Kondo-like phenomenon should occur in more general,
when a localized entity with internal degrees of freedom is embedded
in a conduction electron system and quantum-mechanical exchange
interaction effectively works between local degrees of freedom and
conduction electrons.

In particular, a possibility of Kondo phenomenon with phonon origin
has been pointed out by Jun Kondo himself in a conduction electron system
coupled with a local double-well potential.\cite{Kondo1,Kondo2}
The Kondo effect in such a two-level system has been discussed
actively by several groups.
\cite{Vladar1,Vladar2,Vladar3,Yu-Anderson,Matsuura1,Matsuura2}
Recently, the research of the phononic Kondo effect
has revived due to active experimental investigations
on cage-structure compounds in which a guest atom is contained
in a cage composed of relatively light atoms and
it oscillates with large amplitude in a potential with strong anharmonicity.
Such local oscillation with large amplitude is called rattling and
exotic phenomena induced by rattling have attracted much attention
in the research of strongly correlated electron materials with
cage structure.\cite{SkutReview}

Among them, magnetically robust heavy-electron state has been
vigorously investigated both from experimental and
theoretical sides.\cite{Sanada,Yotsuhashi,Hattori1,Hattori2,
Mitsumoto1,Mitsumoto2,Hotta1,Hotta2,Hotta3,Hotta4,Hotta5,Hotta6,
Yashiki1,Yashiki2,Yashiki3,Hattori3,Hotta7,
Fuse1,Fuse2,Fuse3,Fuse4,Fuse5,Fuse6}
The non-magnetic Kondo effect with phonon origin has been discussed
for the purpose to understand peculiar
magnetically robust heavy-electron state.
In these theoretical investigations, anharmonic Holstein phonon
has been frequently discussed, but it is possible to bring rich physics
when we consider Jahn-Teller vibration of rattling in cage materials.
In fact, when Jahn-Teller phonons are dynamically coupled
with orbital degree of freedom of electrons,
the local ground state is known to be a vibronic state
characterized by clockwise and anti-clockwise rotational
mode of Jahn-Teller phonons.\cite{Hotta1,Hotta2,Hotta4}
Such chiral degree of freedom is screened by orbital moments
of conduction electrons and there occurs the release of an entropy
$\log 2$ of the vibronic state, leading to the Kondo-like phenomenon.

Although the effect of a weak anharmonicity has been partly considered
by the present author to confirm the relation between the Kondo temperature
and the static Jahn-Teller energy,\cite{Hotta1}
previous research has been done mainly for the harmonic potential
of Jahn-Teller ion in order to focus on the Kondo effect due to
chiral degree of freedom of vibronic state.
Here we have an interest on the effect of strong anharmonicity
in the potential for Jahn-Teller ion.
It has been well known that three potential minima appears
when cubic anharmonicity is included in the potential
for Jahn-Teller ion.
In particular, it is worth to investigate the competition between
the positional degree of freedom of three potential minima
and the chiral degree of freedom of vibronic state.

In this paper, we numerically analyze the Jahn-Teller-Anderson model
including the coupling between conduction electrons and
anharmonic Jahn-Teller phonons on an impurity site.
For the convenience of the calculation,
we introduce unitary transformations of electron and phonon operators,
which naturally express the rotational mode of Jahn-Teller phonons.
By using a numerical renormalization group method,
we evaluate entropy, specific heat, and several kinds of susceptibilities.
It is found that the magnitude of entropy changes in the order of
$\log 3$, $\log 2$, and $0$, when we decrease a temperature.
The appearance of $\log 3$ entropy originates from
quasi-triple degeneracy of local low-energy states,
composed of vibronic ground states and the first excited state.
The entropy of $\log 2$ is due to the double degeneracy of
the vibronic ground state composed of electron and dynamical
Jahn-Teller vibration.
We discuss the origin of the temperature at which
entropy changes from $\log 3$ to $\log 2$ in addition to
the Kondo temperature $T_{\rm K}$
corresponding to the entropy release of $\log 2$.
We attempt to explain the behavior of $T_{\rm K}$
by introducing the effective $s$-$d$ model from the original Hamiltonian.

The organization of this paper is as follows.
In Sec.~2, we explain the properties of local phonon and
vibronic states.
In particular, we focus on the local vibronic ground state
and the first excited state.
In Sec.~3, we introduce the Jahn-Teller-Anderson model
to discuss the Kondo phenomenon.
We also briefly explain the method used in this paper and
define several kinds of susceptibilities.
In Sec.~4, we exhibit our numerical results and discuss
how an entropy changes with the decrease of a temperature.
Then, we clarify the energy scale for the entropy release
from $\log 3$ to $\log 2$ as well as the Kondo temperature
corresponding to the entropy release of $\log 2$.
We also introduce the effective $s$-$d$ model to discuss
the behavior of $T_{\rm K}$.
Finally, in Sec.~5, we provide a few comments on
related future problems and summarize this paper.
Throughout this paper, we use such units as $\hbar$=$k_{\rm B}$=1.

%%%%%%%%%%%%%%%%%%%%%%%%%%%%%%%%%%%%%%%%%%
%   Local problem
%%%%%%%%%%%%%%%%%%%%%%%%%%%%%%%%%%%%%%%%%%
\section{Local Phonon and Vibronic States}

\subsection{Anharmonic Potential}

First let us consider the Jahn-Teller oscillation in an anharmonic potential.
The Hamiltonian is given by
\begin{equation}
  H_{\rm ph} = \frac{1}{2M}(P_{2}^2+P_{3}^2)+V(Q_{2},Q_{3}),
\end{equation}
where $M$ is the reduced mass of Jahn-Teller oscillator,
$Q_2$ and $Q_3$ denote normal coordinates of $(x^2-y^2)$- and
$(3z^2-r^2)$-type Jahn-Teller oscillation, respectively,
$P_2$ and $P_3$ indicate corresponding canonical momenta,
and $V(Q_{2},Q_{3})$ is the potential for the Jahn-Teller oscillator.
Here we do not use $Q_1$, since it usually indicates
breathing-mode oscillation, which is ignored in this paper.
Then, the potential is given by
\begin{equation}
\begin{split}
   V(Q_2, Q_3) &= A(Q_2^2+Q_3^2)+B(Q_3^3-3 Q_2^2 Q_3) \\
   &+C(Q_2^2+Q_3^2)^2,
\end{split}
\end{equation}
where $A$ indicates the quadratic term of the potential,
while $B$ and $C$ are, respectively, the coefficients
for third- and fourth-order anharmonic terms.
Note that we consider only the anharmonicity
which maintains the cubic symmetry.
Here we consider the case of $A$$>$$0$ and $C$$>$$0$,
while $B$ takes both positive and negative values.

Before proceeding to the quantization,
we briefly explain the properties of the potential.
For the purpose, it is convenient to introduce
the non-dimensional distortion as
$q_2$=$Q_2/\ell$ and $q_3$=$Q_3/\ell$,
where $\ell$ denotes the amplitude of the zero-point oscillation,
given by $\ell$=$1/\sqrt{2M\omega}$ and
$\omega$ is the phonon energy given by $\omega$=$\sqrt{2A/M}$.
With the use of $q_2$ and $q_3$,
we express $V$ in the unit of $\omega$ as
\begin{equation}
 \begin{split}
 V(q_2, q_3) &= \omega \Bigl[
 \frac{1}{4} (q_2^2+q_3^2)
 +\frac{\beta}{3} (q_3^3-3 q_2^2 q_3) \\
 &+\frac{\gamma}{8}(q_2^2+q_3^2)^2 \Bigr],
\end{split}
\end{equation}
where $\beta$ and $\gamma$ are non-dimensional anharmonicity
parameters, defined by
\begin{equation}
 \beta=\frac{3B}{(2M)^{3/2}\omega^{5/2}}, 
 \gamma=\frac{2C}{M^2\omega^3}.
\end{equation}
By introducing further $q$ and $\theta$ through
the relations of $q_3$=$q \cos \theta$ and $q_2$=$q \sin \theta$,
we rewritten $V$ as
\begin{equation}
 \label{eq:pot}
 V(q, \theta) = \omega \Bigl(
  \frac{q^2}{4} +\frac{\beta q^3 \cos 3\theta}{3}
+\frac{\gamma q^4}{8} \Bigr).
\end{equation}
For $|\beta|$$\le$$\sqrt{\gamma}$, there exists a single minimum at $q$=$0$,
while for $|\beta|$$>$$\sqrt{\gamma}$, there appear three minima
for $q$$\ne$$0$ in addition to the shallow minimum at $q$=$0$.

%%%%%%%%%%%%%% Fig. 1 %%%%%%%%%%%%%%%%
\begin{figure}[t]
\centering
\includegraphics[width=6.7truecm]{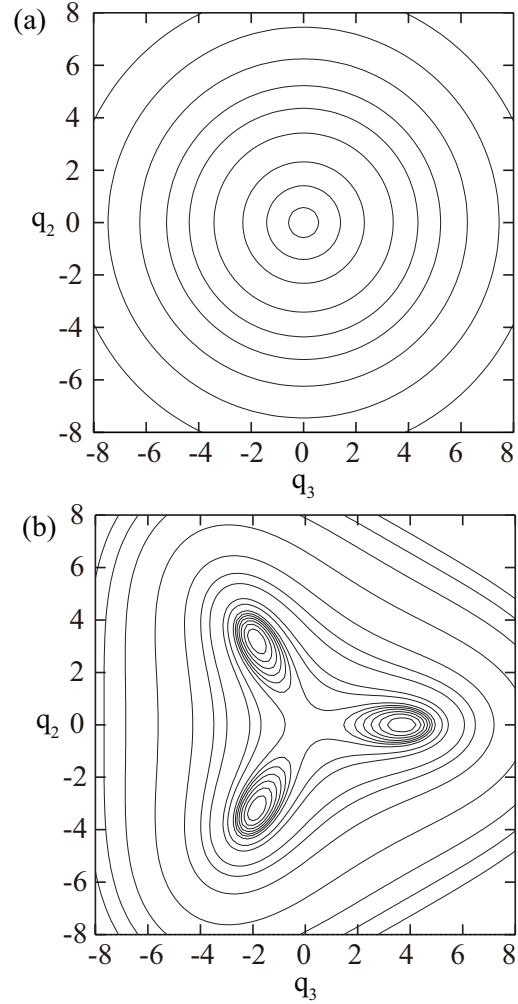}
\caption{Contour plot of $V(q_2, q_3)/\omega$ on the $q_3$-$q_2$ plane
for (a) $\beta$=$0$ and (b) $\beta$=$-2$.
Here we set $\gamma$=$1$.
For the case of (a), we draw the contour curves for
$V(q_2, q_3)/\omega$=$0.1$, $1$, $5$, $20$, $50$, $100$, $200$, $400$,
and $800$, while for (b), we plot the contours for
$V(q_2, q_3)/\omega$=$-6$, $-5$, $-4$, $-3$, $-2$, $-1$, $0.5$, $5$, $10$,
$30$, $50$, $100$, $250$, $500$, and $750$.
}
\end{figure}
%%%%%%%%%%%%%%%%%%%%%%%%%%%%%%%%%

In Fig.~1, we show the contour plot of the potential $V$
on the $q_3$-$q_2$ plane for $\gamma$=$1$.
Figure 1(a) denotes the case of $\beta$=$0$.
The fourth-order anharmonicity is included, but
the potential minimum still appears at the origin.
The increase of the potential for large $q_2$ and $q_3$
becomes rapid, but there is no significant difference
in comparison with the harmonic case
of $\beta$=$\gamma$=$0$.
In Fig.~1(b), we show the case of $\beta$=$-2$,
in which we clearly observe the cubic symmetry in the $q_3$-$q_2$ plane
due to the third-order anharmonicity, instead of the rotational symmetry
for the case of $\beta$=$0$.
At the present case with $\beta$=$-2$, we find three minima
along the directions of $\theta$=$0$, $2\pi/3$, and $4\pi/3$,
corresponding to $(3z^2-r^2)$-, $(3x^2-r^2)$-, and
$(3y^2-r^2)$-type Jahn-Teller distortions, respectively.

\subsection{Phonon State}

In order to discuss the local phonon states,
it is necessary to perform the quantization procedure
through the relations of $q_2$=$a_2+a_2^{\dag}$ and
$q_3$=$a_3+a_3^{\dag}$, where $a_2$ and $a_3$ are
annihilation operators of phonons for Jahn-Teller oscillations.
Then, the Hamiltonian is rewritten as
\begin{equation}
\begin{split}
  H_{\rm ph} &= \omega (a_2^{\dag}a_2+a_3^{\dag}a_3+1) \\ 
  &+ (\omega\beta/3)[(a_3+a_3^{\dag})^3
                               -3(a_2+a_2^{\dag})^2(a_3+a_3^{\dag})] \\
  &+ (\omega\gamma/8)[(a_2+a_2^{\dag})^2+(a_3+a_3^{\dag})^2]^2.
\end{split}
\end{equation}
It is possible to diagonalize this Hamiltonian $H_{\rm ph}$,
but in order to clarify the conserved quantities in the dynamical
Jahn-Teller oscillation, it is useful to introduce
the following transformation for phonon operators:\cite{Takada}
\begin{equation}
  a_{\pm}=(a_3 \pm {\rm i}a_2)/\sqrt{2},
\end{equation}
where the sign in this equation is intuitively understood
as the rotational direction in the potential.
As shown below, in these bases, it is easy to classify
the eigenstates by the symmetry of the Hamiltonian.
In a numerical renormalization group method,
it is quite useful to label the eigenstates with the use of
conserved quantities to keep the precision with the limited
number of the state.

With the use of these operators, after some algebraic calculations,
the Hamiltonian is rewritten as
\begin{equation}
  \begin{split}
  H_{\rm ph} &= \omega (a_{+}^{\dag}a_{+}+a_{-}^{\dag}a_{-}+1) \\
  &+\frac{\omega \beta}{3} [(a_{+} + a_{-}^{\dag})^3
   +(a_{-} + a_{+}^{\dag})^3] \\
  &+\frac{\omega \gamma}{2} (a_{+}^{\dag}a_{+}+a_{-}^{\dag}a_{-}+1
   +a_{+}^{\dag}a_{-}^{\dag}+a_{+}a_{-})^2.
  \end{split}
\end{equation}
When we diagonalize this Hamiltonian, we use the phonon basis expressed by
$|L; n \rangle$, where $L$ has a physical meaning of angular momentum,
while $n$ indicates the phonon number.
Explicitly we define them as
\begin{equation}
  \label{eq:phbasis}
  |L; n \rangle = \left \{
  \begin{array}{ll}
    |L+n, n \rangle & L \ge 0 \\
    |n, n+|L| \rangle & L <0,
  \end{array}
\right.
\end{equation}
where the phonon basis $|n_+,n_- \rangle$ is defined by
\begin{equation}
  |n_+,n_- \rangle=
  \frac{1}{\sqrt{n_+! n_-!}}
  (a_+^{\dag})^{n_+}(a_-^{\dag})^{n_-}|0\rangle.
\end{equation}

For the case of $\beta$=$0$, the phonon state is labelled by
the angular momentum $L$, of which operator is given by
\begin{equation}
  L=a_{+}^{\dag}a_{+}-a_{-}^{\dag}a_{-}.
\end{equation}
Note that $L$ commutes with $H_{\rm ph}$ for $\beta$=$0$.
Thus, when we diagonalize the Hamiltonian for $\beta$=$0$,
we prepare the phonon basis for a fixed value of $L$,
since the states with different $L$ are not mixed.
The physical meaning of $L$ is clear.
When the potential has continuous rotational symmetry for $\beta$=$0$,
the angular momentum should be the conserved quantity in
the quantum mechanics.

However, when we include the effect of $\beta$, namely,
cubic anharmonicity, the situation is changed.
As easily understood from eq.~(\ref{eq:pot}),
there occurs a trigonal term in the potential and in such a case,
$L$ is not the good quantum number,
but there still exists conserved quantity concerning $L$.
In order to clarify this point, we express $L$ as
\begin{equation}
  \label{eq:qn3ph}
  L=3\ell+L_0,
\end{equation}
where $L_0$ takes the values of $0$ and $\pm 1$.
We find that $L_0$ becomes the conserved quantity for $\beta$$\ne$$0$.
Thus, the local phonon state is expressed as
\begin{equation}
  |\Phi^{(0)}_{k,L_0} \rangle
  =\sum_{\ell,n} P_{\ell,n}^{(k,L_0)} |3\ell+L_0; n\rangle,
\end{equation}
where $|\Phi^{(N)}_{k,L_0} \rangle$ denotes the $k$-th eigenstate
of $H_{\rm ph}$ characterized by quantum number $L_0$,
$N$ denotes electron number, and $P$ is the coefficient of the eigenstates.
The corresponding eigenenergy is expressed by $E^{(N)}_{k,L_0}$.

Note that for the case of $L_0$=$0$, there exists extra conserved quantity,
which is a parity concerning the change of $\ell \rightarrow -\ell$.
It is understood from the fact that the bonding and anti-bonding states of
$|3\ell;n\rangle$ and $|-3\ell;n\rangle$ are not mixed with each other.
Then, the parity for the bonding (even) or anti-bonding (odd) state is
another good quantum number.
When we pursue the properties of phonon states more precisely,
it is important to consider the parity.
However, such discussion is meaningful in the high-temperature region
and we do not mention it anymore in the present paper, since we are
interested in the low-temperature properties of the Kondo phenomena.
The peculiar properties of anharmonic Jahn-Teller vibration
in the high-temperature region will be discussed in a separate paper
in the near future.

\subsection{Vibronic State}

Next we consider the electron-phonon coupled state.
For the purpose, we calculate the eigenstate of $H_{\rm loc}$, given by
\begin{equation}
   H_{\rm loc}= H_{\rm el-ph} + H_{\rm ph}.
\end{equation}
Here $H_{\rm el-ph}$ denotes the electron-phonon coupling term, given by
\begin{equation}
   H_{\rm el-ph} = g (\tau_x Q_2 + \tau_z Q_3),
\end{equation}
where $\tau_x$=$d^{\dag}_a d_b + d^{\dag}_b d_a$,
$\tau_z$=$d^{\dag}_a d_z - d^{\dag}_b d_b$,
$d_{\tau}$ is the annihilation operator of spinless fermion with orbital $\tau$,
$a$ and $b$ correspond to $x^2-y^2$ and $3z^2-r^2$ orbitals, respectively,
and $g$ is the electron-phonon coupling constant.
In accordance with the transformation of phonon operator,
we introduce the pseudo-spin expression for electron operators as
\begin{equation}
  d_{\sigma}=(d_a \pm {\rm i}d_b)/\sqrt{2},
\end{equation}
where $\sigma$ denotes a pseudospin and $\uparrow$ ($\downarrow$)
corresponds to $+$ ($-$) sign.
Then, we obtain
\begin{equation}
  H_{\rm el-ph} = \sqrt{2\alpha}\omega
  [(a_+ + a_-^{\dag}) \sigma_+
  +(a_- + a_+^{\dag}) \sigma _-],
\end{equation}
where $\alpha$ is the non-dimensional electron-phonon coupling constant,
given by $\alpha$=$g^2/(2M\omega^3)$,
$\sigma_{+}$=$d^{\dag}_{\uparrow} d_{\downarrow}$,
and $\sigma_{-}$=$d^{\dag}_{\downarrow} d_{\uparrow}$.
In this paper, we consider the case of half-filling,
in which average electron number at an impurity site is unity.
Thus, the local ground state is found in the one-electron sector of $H_{\rm loc}$.

For the harmonic Jahn-Teller phonons with $\beta$=$\gamma$=$0$,
we find the vibronic ground state with double degeneracy, characterized by
the total angular momentum $J$,\cite{Takada} given by
\begin{equation}
  J=L+\sigma_z/2,
\end{equation}
where $\sigma_z$ denotes the $z$-component of pseudospin, given by
$\sigma_z$=$d^{\dag}_{\uparrow} d_{\uparrow}
$$-$$d^{\dag}_{\downarrow} d_{\downarrow}$.
With the use of $J$, we express the eigenstate as
\begin{equation}
  \begin{split}
  |\Phi^{(1)}_{k,J} \rangle
  = \sum_{n} & [ Q_{n}^{(k,J)}d^{\dag}_{\uparrow}
  |J-1/2; n\rangle \\
  &+R_{n}^{(k,J)}d^{\dag}_{\downarrow}
  |J+1/2; n\rangle ],
  \end{split}
\end{equation}
where $J$ takes half-odd-integer value as $J$=$\pm 1/2$, $\pm 3/2$, $\cdots$.
The double degeneracy in the ground states with $J$=$\pm 1/2$ originates from
clockwise and anti-clockwise rotational directions of vibronic states.

When we include the cubic anharmonicity, $J$ is not the good quantum number,
but there still exists a label of total angular momentum $J_0$=$L_0+1/2$
to specify the eigenstate.
Since $L_0$ takes $0$ and $\pm 1$, $J_0$ becomes $\pm 1/2$ and $\pm 3/2$.
Note, however, that the state with $J_0$=$-3/2$ belongs to the same group as
$J_0$=$3/2$, since $J_0$=$-3/2$ is equal to $J_0$=$3/2$ when we add three,
as understood from eq.~(\ref{eq:qn3ph}).
Then, we obtain three groups for the vibronic states characterized by
$J_0$=$+1/2$, $-1/2$, and $3/2$.
Explicitly, they are expressed as
\begin{equation}
  \begin{split}
  |\Phi^{(1)}_{k,+1/2} \rangle
  =\sum_{\ell,n}
  & [Q_{\ell,n}^{(k,1/2)} d^{\dag}_{\uparrow} |3\ell; n\rangle \\
  & +R_{\ell,n}^{(k,1/2)} d^{\dag}_{\downarrow} |3\ell+1; n\rangle],
\end{split}
\end{equation}
\begin{equation}
  \begin{split}
  |\Phi^{(1)}_{k,-1/2} \rangle
  =\sum_{\ell,n}
  &[Q_{\ell,n}^{(k,-1/2)} d^{\dag}_{\uparrow} |3\ell-1; n\rangle \\
  &+R_{\ell,n}^{(k,-1/2)} d^{\dag}_{\downarrow} |3\ell; n\rangle],
\end{split}
\end{equation}
and
\begin{equation}
  \begin{split}
  |\Phi^{(1)}_{k,3/2} \rangle
  =\sum_{\ell,n}
  &[Q_{\ell,n}^{(k,3/2)} d^{\dag}_{\uparrow} |3\ell+1; n\rangle \\
  &+R_{\ell,n}^{(k,3/2)} d^{\dag}_{\downarrow} |3\ell-1; n\rangle],
\end{split}
\end{equation}
respectively.

%%%%%%%%%% Fig. 2 %%%%%%%%%%%%%%%%
\begin{figure}[t]
\centering
\includegraphics[width=8.4truecm]{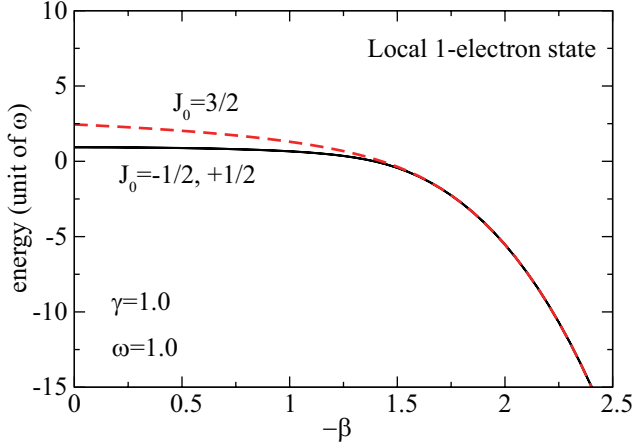}
\caption{
Eigenenergies vs. $-\beta$ for the states of $J_0$=$\pm 1/2$ and $3/2$.
}
\end{figure}
%%%%%%%%%%%%%%%%%%%%%%%%%%%%%

In Fig.~2, we show the corresponding eigenenergies of $E^{(1)}_{0, J_0}$
as functions of $-\beta$ for $\gamma$=$1$.
It is observed that the ground state has still double degeneracy, characterized
by $J_0$=$\pm 1/2$, even when we consider the cubic anharmonicity.
The state with $J_0$=$3/2$ is the excited state, but
$E^{(1)}_{0,3/2}$ seems to be almost equal to $E^{(1)}_{0,\pm 1/2}$
in the region of $-\beta$$>$$1.5$.
The excitation energy becomes exponentially small for large $|\beta|$.
This quasi degeneracy is understood as follows:
The potential minima become deep and the quantum tunneling motion
is suppressed, when $|\beta|$ is increased.
Thus, the energy difference due to rotational motion becomes very small.

For the case of $N$=$2$ with double occupancy, the coupling between
electrons and Jahn-Teller vibration becomes inactive.
Then, the eigenstate is written as
\begin{equation}
  |\Phi^{(2)}_{k,L_0} \rangle
  =\sum_{\ell,n} P_{\ell,n}^{(k,L_0)}
  d^{\dag}_{\uparrow} d^{\dag}_{\downarrow} |3\ell+L_0; n\rangle,
\end{equation}
where the coefficient $P$ is the same as that of 
$|\Phi^{(0)}_{k, L_0} \rangle$.

%%%%%%%%%%%%%%%%%%%%%%%%%%%%%%%%%%%
%   Model and Method
%%%%%%%%%%%%%%%%%%%%%%%%%%%%%%%%%%%
\section{Model and Method}

Now we consider the conduction electron hybridized with localized electrons.
Then, the model is expressed as
\begin{equation}
  H=\sum_{\mib{k}\tau} \varepsilon_{\mib{k}}
    c_{\mib{k}\tau}^{\dag} c_{\mib{k}\tau}
   +\sum_{\mib{k}\tau} (Vc_{\mib{k}\tau}^{\dag}d_{\tau}+{\rm h.c.})
   +H_{\rm loc},
\end{equation}
where $\varepsilon_{\mib{k}}$ is the dispersion of conduction electron,
$c_{\mib{k}\tau}$ is an annihilation operator of conduction electron
with momentum $\mib{k}$ and orbital $\tau$,
and $V$ is the hybridization between conduction and localized electrons.
The energy unit is a half of the conduction bandwidth, $D$,
which is set as unity in the following.

In order to investigate electronic and phononic properties of $H$
at low temperatures, we usually discuss corresponding susceptibilities.
The susceptibility of an arbitrary operator $A$ is expressed by
\begin{equation}
  \label{sus}
  \chi_{A}=\frac{1}{Z} \sum_{n,m}
  \frac{e^{-E_n/T}-e^{-E_m/T}}{E_m-E_n}
  | \langle m | A | n \rangle|^2,
\end{equation}
where $E_n$ is the eigenenergy for the $n$-th eigenstate
$|n\rangle$ of $H$ and $Z$ is the partition function given by
$Z$=$\sum_n e^{-E_n/T}$.

For the evaluation of susceptibilities, here we employ
a numerical renormalization group (NRG) method,\cite{NRG1,NRG2}
in which momentum space is logarithmically discretized
to include efficiently the conduction electrons near the Fermi energy
and the conduction electron states are characterized by ``shell'' labeled by $N$.
The shell of $N$=$0$ denotes an impurity site described by the local Hamiltonian.
The Hamiltonian is transformed into the recursion form as
\begin{eqnarray}
  H_{N+1} = \sqrt{\Lambda} H_N+t_N \sum_{\sigma}
  (c_{N\sigma}^{\dag}c_{N+1\sigma}+c_{N+1\sigma}^{\dag}c_{N\sigma}),
\end{eqnarray}
where $\Lambda$ is a parameter for logarithmic discretization,
$c_{N\sigma}$ denotes the annihilation operator of conduction electron
in the $N$-shell, and $t_N$ indicates ``hopping'' of electron between
$N$- and $(N+1)$-shells, expressed by
\begin{eqnarray}
  t_N=\frac{(1+\Lambda^{-1})(1-\Lambda^{-N-1})}
  {2\sqrt{(1-\Lambda^{-2N-1})(1-\Lambda^{-2N-3})}}.
\end{eqnarray}
The initial term $H_0$ is given by
\begin{equation}
  H_0=\Lambda^{-1/2} [H_{\rm loc}
  +\sum_{\sigma}V(c_{0\sigma}^{\dag}d_{\sigma}
                             +d_{\sigma}^{\dag}c_{0\sigma})].
\end{equation}
We also evaluate entropy and specific heat of localized electron.
In the NRG calculations, a temperature $T$ is defined as
$T$=$\Lambda^{-(N-1)/2}$.
In this paper, we set $\Lambda$=$5$ and we keep $M$=$5000$
low-energy states for each renormalization step.
The phonon basis $|L; n \rangle$ of eq.~(\ref{eq:phbasis})
is truncated at a finite number $N_{\rm ph}$
and a maximum angular momentum $L_{\rm max}$.
In order to check the convergence of the results,
we have performed the numerical calculations for $N_{\rm ph}$
and $L_{\rm max}$ up to 120 and 60, respectively.

%%%%%%%%%%%%%%%%%%%%%%%%%%%%%%%
%   Results
%%%%%%%%%%%%%%%%%%%%%%%%%%%%%%%
\section{Calculated Results}

Now we move on to NRG results of the anharmonic Jahn-Teller-Anderson model.
In Fig.~3(a), we show the results of entropy and specific heat.
In a high-temperature region, an entropy is rapidly decreased
with the decrease of a temperature $T$ and it forms a plateau of $\log 3$
between $10^{-5} < T < 10^{-2}$.
The origin of $\log 3$ entropy is the quasi-degeneracy of local low-energy states.
As mentioned in Sec.~2, for the one-electron case, we find the vibronic
ground states characterized by $J_0$=$\pm 1/2$ with double degeneracy
and the first excited state characterized by $J_0$=$3/2$.
As already shown in Fig.~2, we always obtain the vibronic ground states,
but the excitation energy $\Delta E$ becomes exponentially small
for large $|\beta|$ such as $\beta$=$-2$.
Thus, we observe the $\log 3$ plateau in the entropy as a function of $T$.

%%%%%%%%%% Fig. 3 %%%%%%%%%%
\begin{figure}[t]
\centering
\includegraphics[width=8.4truecm]{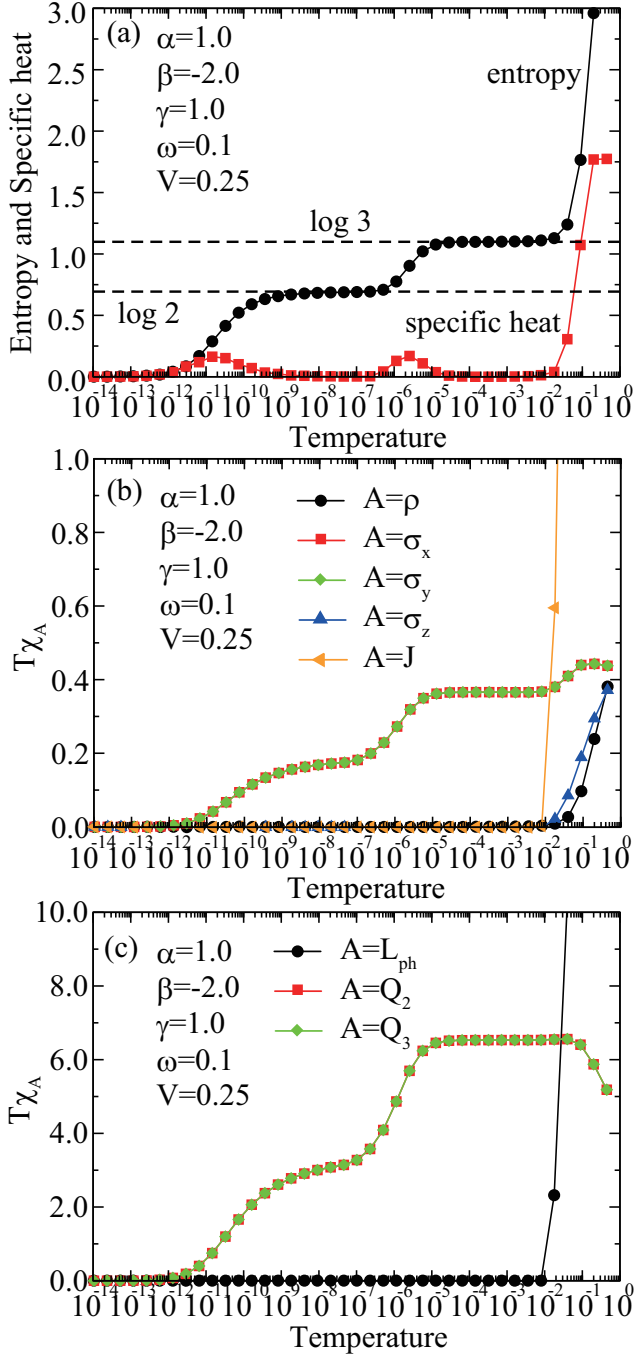}
\caption{
(a) Entropy and specific heat for $\omega$=$0.1$, $\alpha$=$1$,
$\beta$=$-2$, and $\gamma$=$1$.
(a) Susceptibilities for charge, pseudospin, and total angular momentum
for the same parameters as in (a).
(c) Susceptibilities for phonon rotational moment and Jahn-Teller vibration
$Q_2$ and $Q_3$ for the same parameters as in (a).
}
\end{figure}
%%%%%%%%%%%%%%%%%%%%%%%

As easily understood from the above explanation, the quasi-triple degeneracy
should be lifted around at $T$=$\Delta E$, where $\Delta E$ denotes
the first excitation energy among local low-energy states.
In fact, we observe a clear peak in the specific heat at $T$$\approx$$\Delta E$,
since the entropy is changed from $\log 3$ to $\log 2$.
The $\log 2$ entropy originates from the double degeneracy
in the local vibronic states with $J_0$=$\pm 1/2$,
corresponding to clockwise and anti-clockwise rotational directions.
Around at a temperature where the rotational moment is screened by
orbital moments of conduction electrons, the entropy of $\log 2$ is eventually
released and a peak is formed in the specific heat.
This peak naturally defines a characteristic temperature,
which is called here the Kondo temperature $T_{\rm K}$.

In order to confirm the relevant quantity of the present Kondo phenomenon,
we investigate several kinds of susceptibilities.
As shown in Fig.~3(b), we observe that susceptibilities for charge ${\rho}$,
$z$-component of pseudospin ${\sigma_z}$, and total angular momentum
$J$ are rapidly suppressed in the high-temperature region.
On the other hand, we observe that $\chi_{\sigma_x}$=$\chi_{\sigma_y}$
and they show characteristic behavior at temperatures
where entropy is released.
Note here that $\sigma_x$=$\sigma_{+}+\sigma_{-}$ and
$\sigma_y$=$-{\rm i}(\sigma_{+}-\sigma_{-})$ in the chiral bases,
but they are transformed to $\tau_x$ and $\tau_z$ in the original Hamiltonian.
Since Jahn-Teller distortions $Q_2$ and $Q_3$ are coupled with $\tau_x$ and
$\tau_z$, it is quite natural to find the similar characteristic behavior
for the susceptibilities of $Q_2$ and $Q_3$, as shown in Fig.~3(c).
On the other hand, the susceptibility for phonon rotational moment $L$
is rapidly decreased in the high-temperature region.

For the case without anharmonicity, we have clearly observed that
the total angular momentum $J$,
composed of pseudo spin and phonon angular momenta,
is screened to form the singlet ground state of $J$=$0$.\cite{Hotta1,Hotta2}
On the other hand, in the present case with strong anharmonicity,
$J$ itself is no longer a good quantum number.
However, the rotational direction, clockwise or anti-clockwise, is still
a relevant degree of freedom, as understood from the fact that
the local ground states are characterized by $J_0$=$\pm 1/2$.
In other words, orbital (quadrupole) degree of freedom plays
a crucial role in the present Kondo phenomenon.
Concerning the viewpoint of the actual detection of this Kondo phenomenon,
quadrupole susceptibilities for $Q_2$ and $Q_3$ are quite important,
since they are closely related to the elastic constant observed
by ultrasound experiments.

In Fig.~4(a), we plot the characteristic temperatures (solid symbols)
which form peaks in the specific heat for the case of
$\omega$=$0.1$, $\alpha$=$1$, and $\gamma$=$1$.
Note that in this case, in the region of $|\beta|<1.5$,
we do not find any peak temperatures in the specific heat,
since the entropy is rapidly decreased to zero in the high-temperature region.
For $|\beta|>1.5$, we observe two peaks in the specific heat.
As mentioned above, the higher one is characterized by the
local excitation energy $\Delta E$, while the lower one is
considered as the Kondo temperature $T_{\rm K}$.
In Fig.~4(a), the solid circles are well fitted by solid curve of $\Delta E$,
which is obtained by the diagonalization of $H_{\rm loc}$.
We observe small deviations between solid circles and solid curve,
but they are due to the effect of discrete temperature of the NRG calculation.

Note that in the case of $\omega$=$0.1$,
the situation is in the adiabatic region and the dynamical Jahn-Teller
effect is not so significant.
Thus, it is difficult to obtain the formula of the Kondo temperature
concerning the rotational degree of freedom of the vibronic
state as a result of the dynamical Jahn-Teller effect.
However, in the previous research, we have found that
in the adiabatic region, $\log T_K$ is given by the form of
$a E_{\rm JT}+b$ with the use of the static Jahn-Teller energy
$E_{\rm JT}$ and appropriate parameters $a$ and $b$.\cite{Hotta1}
When we include the anharmonicity, we have found the relation of
$E_{\rm JT}$$\propto$$|\beta|$ for fixed values of $\alpha$
and $\gamma$.\cite{Hotta1}
In fact, as shown in Fig.~4(a), we find that $\log T_{\rm K}$ is
in proportion to $|\beta|$ except for the value of constant shift $b$.
This result suggests that the static Jahn-Teller energy should be
relevant to the Kondo temperature in the adiabatic region.

Now we increase the value of $\omega$
in order to check the effect of adiabaticity.
We do not show the NRG results of entropy, specific heat, and
several kinds of susceptibilities, since we have obtained essentially
the same results as Figs.~3 for the case of $\omega$=$0.5$.
in Fig.~4(b), we plot the temperatures of the peak positions
in the specific heat for $\omega$=$0.5$.
Again we find that the higher peak temperature is well fitted
by $\Delta E$ for the case of $\omega$=$0.5$.
Note that the magnitude of $\Delta E$ becomes totally larger than that in
Fig.~4(a), since $\omega$ determines the energy scale of $H_{\rm loc}$.
The lower peak is the Kondo temperature and in the present case,
it is possible to fit the NRG results by a single curve of the Kondo temperature
$T_{\rm K}$, which is obtained from the effective $s$-$d$ model,
discussed below.

%%%%%%%%%%% Fig. 4 %%%%%%%%%%%%%%%
\begin{figure}[t]
\centering
\includegraphics[width=8.4truecm]{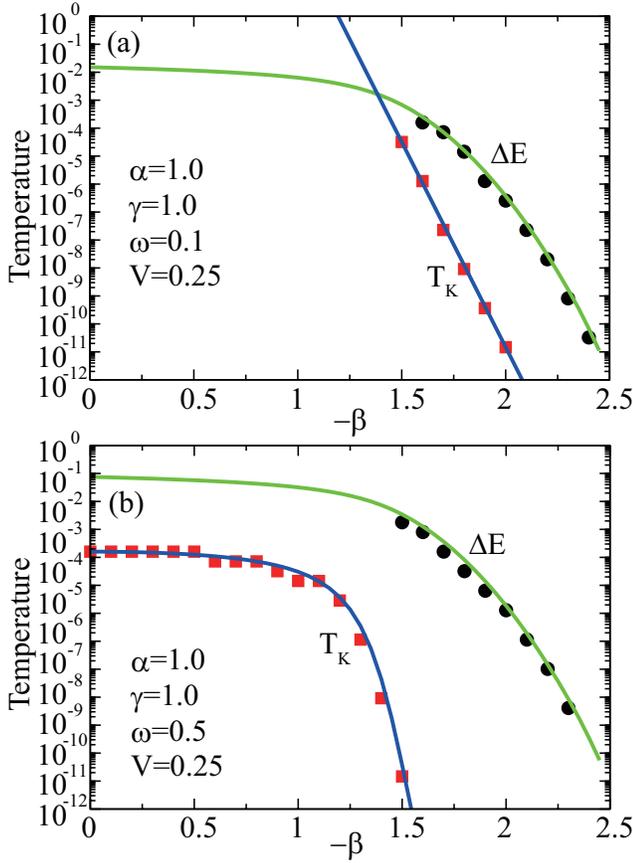}
\caption{
Phase diagram for the Kondo effect of a Jahn-Teller vibration
in a cubic anharmonic potential for (a) $\omega$=$0.1$ and
(b) $\omega$=$0.5$ with $\alpha$=$1$ and $\gamma$=$1$.
Solid symbols are obtained from the peak positions in the NRG results.
Solid curves are drawn by the local energy $\Delta E$ and
the fitting curves for $T_{\rm K}$.
}
\end{figure}
%%%%%%%%%%%%%%%%%%%%%%%%%%%%%%%

Let us explain the effective $s$-$d$ model of
the Jahn-Teller-Anderson Hamiltonian.
First we distinguish the local degenerate ground states of
$H_{\rm loc}$ as $|\pm \rangle$=$|\Phi^{(1)}_{0,\pm 1/2} \rangle$
and the ground-state energy is given by $E^{(1)}_{0,\pm 1/2}$.
After some algebraic calculations concerning the second-order perturbation
in terms of the hybridization, we obtain the effective $s$-$d$ model as
\begin{equation}
  \begin{split}
  H_{\rm eff} &= \sum_{\mib{k}\sigma} \varepsilon_{\mib{k}}
  c_{\mib{k}\sigma}^{\dag} c_{\mib{k}\sigma}
  +\sum_{{\mib k},{\mib k'}}
  [J_z(c_{\mib{k}\uparrow}^{\dag} c_{\mib{k'}\uparrow}
  -c_{\mib{k}\downarrow}^{\dag} c_{\mib{k'}\downarrow})S_z \\
  &+J_{\bot}(c_{\mib{k}\downarrow}^{\dag} c_{\mib{k'}\uparrow}S_+
  +c_{\mib{k}\uparrow}^{\dag} c_{\mib{k'}\downarrow}S_-)],
  \end{split}
\end{equation}
where $S_z$=$(|+ \rangle \langle +|-|-\rangle \langle -|)/2$,
$S_{+}$=$|+ \rangle \langle - |$, $S_{-}$=$|- \rangle \langle + |$,
and the exchange interactions are given by
\begin{equation}
   J_{\bot} = \sum_{k}
   \frac{2V^2 a_{k}^2}{E^{(0)}_{k,0} - E^{(1)}_{0,1/2}},
\end{equation}
and
\begin{equation}
   J_z = \sum_{k}
   \frac{2V^2 (a_{k}^2-b_{k}^2)}{E^{(0)}_{k,0} - E^{(1)}_{0,1/2}}.
\end{equation}
Here $a_k$ and $b_k$ are given by
\begin{equation}
  a_{k}=\sum_{n,\ell} Q^{(0,1/2)}_{n,\ell} P^{(k,0)}_{n,\ell},
\end{equation}
and
\begin{equation}
  b_{k}=\sum_{n,\ell} R^{(0,1/2)}_{n,\ell} P^{(k,1)}_{n,\ell},
\end{equation}
respectively.
Except for the definitions of $J_{\bot}$ and $J_z$, the model is essentially
the same as that for the harmonic Jahn-Teller phonons.\cite{Hotta2}

Here we note that $J_z$ is always smaller than $J_{\bot}$
and it is decreased when $|\beta|$ is increased,
leading to the rapid suppression of $T_{\rm K}$
as a function of $|\beta|$, as observed in Fig.~4(b).
The effect of anharmonicity appears in the suppression of
the $z$-component of the exchange interaction.
This is consistent with the relation of
$\chi_{\sigma_x}$=$\chi_{\sigma_y}$$>$$\chi_{\sigma_z}$
for $T$$>$$T_{\rm K}$, since Jahn-Teller phonons are coupled
with $\sigma_{+}$ and $\sigma_{-}$.
The directions of the rotational Jahn-Teller modes
are easily converted for high-energy Jahn-Teller phonons.

For the $s$-$d$ model with anisotropic exchange interaction,
the explicit expression for the binding energy ${\tilde E}$
has been already obtained.\cite{Shiba}
When we define the Kondo temperature $T_{\rm K}$ as
$T_{\rm K}$=$-{\tilde E}$, we express $T_{\rm K}$ as
\begin{equation}
  \label{Tk}
  T_{\rm K}=D {\rm exp} \Biggl[
  \frac{-1}{2\rho_0\sqrt{J_{\bot}^2-J_z^2}}
  \tan^{-1}\Bigl(\frac{\sqrt{J_{\bot}^2-J_z^2}}{J_z} \Bigr) \Biggr].
\end{equation}
In Fig.~4(b), numerical results are shown by solid squares and
the curve on the solid square denotes $T_{\rm K}$.
We find that the numerical results agree well with the analytic curve of
$T_{\rm K}$ of the $s$-$d$ model with the anisotropic exchange interaction.

%%%%%%%%%%%%%%%%%%%%%%%%%%%%%%%%
%   Discussion and Summary
%%%%%%%%%%%%%%%%%%%%%%%%%%%%%%%%
\section{Discussion and Summary}

In this paper, we have discussed the Kondo phenomena due to
the Jahn-Teller ion vibrating in the cubic anharmonic potential
by using the numerical renormalization group method.
Concerning the magnitude of the anharmonicity,
we have considered the region of $|\beta|$$<$$2.5$ in this paper,
but it is possible, in principle,
to perform the numerical calculations
even for the region of $|\beta|$$>$$2.5$.
However, it is necessary to pay due attention to the reliability
of the numerical results, since the Kondo temperature is expected
to be extremely low in such a case.
In fact, from the viewpoint of the limitation of the numerical calculations,
it is difficult to obtain the reliable results in the present calculation,
when the Kondo temperature becomes the order of $10^{-12}$.
Thus, for the case with very strong anharmonicity,
it is difficult to confirm the results by actual calculations,
but we expect that the curves of $\Delta E$ and $T_{\rm K}$
in Fig.~4(b) are simply extended to the lower-temperature side.
Namely, we deduce that the picture of the two-step entropy release is
invariant even for large $|\beta|$.

Throughout this paper, we have ignored the Coulomb interaction.
As mentioned previously in the research of orbital ordering phenomena
in manganites,\cite{HottaMn1,HottaMn2}
in the present spinless model, the inter-orbital Coulomb interaction
effectively enhances the static Jahn-Teller energy.
This is quite natural, since the Jahn-Teller distortion occurs,
only when one electron is included at an impurity site
and orbital degree of freedom becomes active.
In fact, we simply observe the suppression of $T_{\rm K}$,
when we include the inter-orbital Coulomb interaction in the present model.
Thus, we have not considered the Coulomb interaction in this paper.

However, the situation will be changed,
if we consider explicitly the spin degree of freedom.
In this paper, in order to focus the Kondo effect due to
anharmonic Jahn-Teller vibrations, we have considered only
the spinless model, but we are also interested in the competition
between spin and orbital degrees of freedoms.
On the basis of this viewpoint, recently, we have analyzed
the two-orbital spinful Anderson model including both Coulomb interactions
and the coupling between electrons and Jahn-Teller phonons
on an impurity site.\cite{Fuse4}
A local problem has been analyzed in detail for the confirmation of
a spin-vibronic quartet ground state, which is characterized by
the direct product of spin and vibronic degrees of freedom.
Then, the hybridization term has been included and
the model has been analyzed with the use of the NRG method.
From the evaluation of entropy, specific heat,
and several kinds of susceptibilities,
it has been found that spin and total angular momenta are simultaneously
screened by conduction electrons, leading to the Kondo phenomenon
due to the entropy release of $\log 4$ of the local spin-vibronic state.
It is one of future problems to clarify the effect of cubic anharmonicity
on the Kondo effect due to the spin-vibronic state.

As mentioned in the last paragraph of Sec.~2.2,
we have not discussed in detail the quantum phonon states for the case
of $N$=$0$.
However, it is an interesting problem to clarify the properties of eigenstates
of $H_{\rm ph}$.
Since our interests have focused on the low-temperature
properties of the Kondo phenomena in this paper,
we have not remarked the peculiar properties of anharmonic Jahn-Teller
vibration in the high-temperature region.
This is another future problem.

In summary, we have clarified the Kondo effect
in the Jahn-Teller-Anderson model with cubic anharmonicity.
We have found the $\log 3$ plateau in the entropy
due to quasi-triple degeneracy in the low-energy states
including vibronic ground states.
With the further decrease of temperature, we have observed
the region of $\log 2$ plateau due to the vibronic state
with rotational degree of freedom.
The rotational moment of the vibronic state has been found to
be suppressed by the screening of orbital moments of
conduction electrons, leading to the Kondo effect.
It has been shown that $T_{\rm K}$ is well explained by
the effective $s$-$d$ model with anisotropic exchange interactions.

%%%%%%%%%%%%%%%%%%%%%%%%%%%%%
%   Acknowledgement
%%%%%%%%%%%%%%%%%%%%%%%%%%%%%
\section*{Acknowledgement}

The author is grateful to K. Ueda for fruitful discussions
concerning the Kondo phenomena.
He also thanks Y. Aoki, T. D. Matsuda, R. Higashinaka, and A. Shudo
for discussions on rattling in cage-structure materials.
This work has been supported by a Grant-in-Aid for Scientific Research (C)
(No. 24540379) of Japan Society for the Promotion of Science.
The computation in this work has been partly done using the facilities of the
Supercomputer Center of Institute for Solid State Physics, University of Tokyo.

%%%%%%%%%%%%%%%%%%%%%%%%%%%%
%   References
%%%%%%%%%%%%%%%%%%%%%%%%%%%%


\begin{thebibliography}{99}

\bibitem{ICHE2010}
{\it Proc. Int. Conf. Heavy Electrons (ICHE2010)},
J. Phys. Soc. Jpn. {\bf 80} (2010) Suppl. A.

\bibitem{SCES2013}
{\it Advances in Physics of Strongly Correlated Electron Systems},
J. Phys. Soc. Jpn. {\bf 83} (2014) 061001-061019.

\bibitem{Kondo40}
Kondo effect and its related phenomena have been reviewed
in J. Phys. Soc. Jpn. {\bf 74} (2005) 1-238.

\bibitem{Kondo1}
J. Kondo: Physica B+C {\bf 84} (1976) 40.

\bibitem{Kondo2}
J. Kondo: Physica B+C {\bf 84} (1976) 207.

\bibitem{Vladar1}
K. Vlad\'ar and A. Zawadowski:
Phys. Rev. B {\bf 28} (1983) 1564.

\bibitem{Vladar2}
K. Vlad\'ar and A. Zawadowski:
Phys. Rev. B {\bf 28} (1983) 1582.

\bibitem{Vladar3}
K. Vlad\'ar and A. Zawadowski:
Phys. Rev. B {\bf 28} (1983) 1596.

\bibitem{Yu-Anderson}
C. C. Yu and P. W. Anderson:
Phys. Rev. B {\bf 29} (1984) 6165.

\bibitem{Matsuura1}
T. Matsuura and K. Miyake:
J. Phys. Soc. Jpn. {\bf 55} (1986) 29.

\bibitem{Matsuura2}
T. Matsuura and K. Miyake:
J. Phys. Soc. Jpn. {\bf 55} (1986) 610.

\bibitem{SkutReview}
H. Sato, H, Sugawara, Y. Aoki, and H. Harima:
{\it Handbook of Magnetic Materials} Volume 18,
ed. K. H. J. Buschow, pp. 1-110, Elsevier, Amsterdam, 2009.

\bibitem{Sanada}
S. Sanada, Y. Aoki, H. Aoki, A. Tsuchiya, D. Kikuchi, H. Sugawara,
and H. Sato: J. Phys. Soc. Jpn. {\bf 74} (2005) 246.

\bibitem{Yotsuhashi}
S. Yotsuhashi, M. Kojima, H. Kusunose, and K. Miyake:
J. Phys. Soc. Jpn. {\bf 74} (2005) 49.

\bibitem{Hattori1}
K. Hattori, Y. Hirayama, and K. Miyake:
J. Phys. Soc. Jpn. {\bf 74} (2005) 3306.

\bibitem{Hattori2}
K. Hattori, Y. Hirayama, and K. Miyake:
J. Phys. Soc. Jpn. {\bf 75} (2006) Suppl. 238.

\bibitem{Mitsumoto1}
K. Mitsumoto and Y. {\=O}no:
Physica B {\bf 403} (2008) 859.

\bibitem{Mitsumoto2}
K. Mitsumoto and Y. {\=O}no:
J. Phys. Soc. Jpn {\bf 79} (2010) 054707.

\bibitem{Hotta1}
T. Hotta: Phys. Rev. Lett. {\bf 96} (2006) 197201.

\bibitem{Hotta2}
T. Hotta: J. Phys. Soc. Jpn. {\bf 76} (2007) 023705.

\bibitem{Hotta3}
T. Hotta: J. Phys. Soc. Jpn. {\bf 76} (2007) 084702.

\bibitem{Hotta4}
T. Hotta: Physica B {\bf 403} (2008) 1371.

\bibitem{Hotta5}
T. Hotta: J. Phys. Soc. Jpn. {\bf 77} (2008) 103711.

\bibitem{Hotta6}
T. Hotta: J. Phys. Soc. Jpn. {\bf 78} (2009) 073707.

\bibitem{Yashiki1}
S. Yashiki, S. Kirino, and K. Ueda:
J. Phys. Soc. Jpn. {\bf 79} (2010) 093707.

\bibitem{Yashiki2}
S. Yashiki, S. Kirino, K. Hattori, and K. Ueda:
J. Phys. Soc. Jpn. {\bf 80} (2011) 064701.

\bibitem{Yashiki3}
S. Yashiki and K. Ueda:
J. Phys. Soc. Jpn. {\bf 80} (2011) 084717.

\bibitem{Hattori3}
K. Hattori: Phys. Rev. B {\bf 85} (2012) 214411.

\bibitem{Hotta7}
T. Hotta and K. Ueda:
Phys. Rev. Lett. {\bf 108} (2012) 247214.

\bibitem{Fuse1}
T. Fuse and Y.\=Ono:
J. Phys. Soc. Jpn. {\bf 79} (2010) 093702.

\bibitem{Fuse2}
T. Fuse and Y.\=Ono:
J. Phys. Soc. Jpn. {\bf 80} (2011) SA136.

\bibitem{Fuse3}
T. Fuse, Y. \=Ono, and T. Hotta:
J. Phys. Soc. Jpn. {\bf 81} (2012) 044701.

\bibitem{Fuse4}
T. Fuse and T. Hotta:
J. Phys.: Conf. Ser. {\bf 428} (2013) 012013.

\bibitem{Fuse5}
T. Fuse and T. Hotta:
J. Korean Phys. Soc. {\bf 62} (2013) 1874.

\bibitem{Fuse6}
T. Fuse and T. Hotta:
To appear in the Proceedings of SCES2013.

\bibitem{Takada}
Y. Takada: Phys. Rev. B {\bf 61} (2000) 8631.

\bibitem{NRG1}
K. G. Wilson: Rev. Mod. Phys. {\bf 47} (1975) 773. 

\bibitem{NRG2}
H. R. Krishna-murthy, J. W. Wilkins, and K. G. Wilson,
Phys. Rev. B {\bf 21} (1980) 1003.

\bibitem{Shiba}
H. Shiba: Prog. Theor. Phys. {\bf 43} (1970) 601.

\bibitem{HottaMn1}
E. Dagotto, T. Hotta, and A. Moreo:
Phys. Rep. {\bf 344} (2001) 1.

\bibitem{HottaMn2}
T. Hotta: Rep. Prog. Phys. {\bf 69} (2006) 2061.

\end{thebibliography}
\end{document}